\documentclass[twocolumn,aps,superscriptaddress,showpacs,nofootinbib,floatfix]{revtex4-1}
\usepackage{epsfig,bm,feynmf}
\usepackage{graphics}
\usepackage{amsmath}
\usepackage[normalem]{ulem}  % \sout{old text} for strikeout
\usepackage[dvips]{color} % For blue in-text comments and additions

\renewcommand\sout{\bgroup \color{red} \ULdepth=-.5ex \ULset}
\begin{document}

%%%%%%%%%%%%%%%%%%%%% Title %%%%%%%%%%%%%%%%%%%%%%

\title{$J/\psi$ near $T_c$}

%%%%%%%%%%%%%%%%%%%% Authors %%%%%%%%%%%%%%%%%%%%%

\author{Taesoo Song}\email{t.song@gsi.de}
\affiliation{GSI Helmholtzzentrum f\"{u}r Schwerionenforschung GmbH, Planckstrasse 1, 64291 Darmstadt, Germany}

\author{Philipp Gubler}%
\email{gubler@post.j-parc.jp}
\affiliation{Advanced Science Research Center, Japan Atomic Energy Agency, Tokai, Ibaraki 319-1195, Japan}

\author{Juhee Hong}%
\email{juhehong@yonsei.ac.kr}
\affiliation{Department of Physics and Institute of Physics and Applied Physics, Yonsei University, Seoul 03722, Korea}

\author{Su Houng Lee}%
\email{suhoung@yonsei.ac.kr}
\affiliation{Department of Physics and Institute of Physics and Applied Physics, Yonsei University, Seoul 03722, Korea}

\author{Kenji Morita}%
\email{morita.kenji@qst.go.jp}
\affiliation{RIKEN Nishina Center, Wako 351-0198, Japan}
\affiliation{National Institutes for Quantum and Radiological Science and Technology, Rokkasho Fusion Institute, Rokkasho, Aomori, 039-3212, Japan}

%%%%%%%%%%%%%%%%%%%% Abstract %%%%%%%%%%%%%%%%%%%%%

\begin{abstract}
 We calculate the mass shift and thermal decay width of the
 $J/\psi$ near the QCD transition temperature $T_c$ by imposing
 two independent constraints on these variables that can be obtained
 first by solving the Schr\"odinger equation and second from the QCD sum
 rule approach. While the real part of the potential is determined by comparing the
 QCD sum rule result for charmonium and the D meson to
 that from the potential model result,  the imaginary
 potential is taken to be proportional to the perturbative form
 multiplied by a constant factor, which in turn can be determined
by applying the two independent constraints. The
 result shows that the binding energy  and the thermal
 width becomes similar in magnitude at around $T=1.09T_c$, above which the sum rule analysis
also becomes unstable, strongly suggesting that the $J/\psi$
 will melt slightly above $T_c$.
\end{abstract}

%\pacs{25.75.Nq, 25.75.Ld}
%\keywords{}

\maketitle

{\it Introduction:}
Quarkonium suppression in  relativistic heavy-ion collisions has been the subject of numerous theoretical and experimental studies since the pioneering work of Matsui and Satz~\cite{Matsui:1986dk}, who proposed the phenomenon  as a signature of quark-gluon plasma formation in the early stages of the collisions.
While the $J/\psi$ is expected to melt away at sufficiently high
temperature above the quark-hadron transition temperature $T_c$,
its exact melting point,  thermal width and mass shift above
the transition temperature have been the focal point of intense
theoretical studies often with varying
results~\cite{Hashimoto:1986nn,Hansson:1987uk,Asakawa:2003re,Kim:2018yhk}.
These properties are essential inputs to quantitatively understand the
charmonium suppression in heavy ion collisions (see, for instance, the
recent review in Ref.\,\cite{Rothkopf:2019ipj}) and are also related to
the transport coefficients for heavy quarks at high temperatures~\cite{Brambilla:2019tpt}.

From a potential model point of view, obtaining the correct mass shift
and width is respectively related to identifying the correct real and
imaginary parts of the potential to be used for the heavy quark system
near $T_c$~\cite{Brambilla:2008cx,Kaczmarek:2002mc,Satz:2015jsa}.
Employing the free-energy potential, the $J/\psi$
dissolves around 1.1 $T_c$, whereas it survives up to a higher
temperatures when the internal energy is used
\cite{Wong:2004zr,Kaczmarek:2002mc}.

As for the imaginary part of the potential, even though the high temperature limit can be calculated using the hard-thermal-loop (HTL) resummed perturbation theory \cite{Laine:2006ns}, applying the same formula near $T_c$ might be problematic.
Lattice QCD  will in all likelihood eventually be used to calculate the complex potential~\cite{Rothkopf:2011db} and the corresponding properties of quarkonium near $T_c$, but the present uncertainties in the imaginary potential appear to be still large~\cite{Petreczky:2018xuh}.

Some of us have previously used QCD sum rules to calculate the
mass shift of $J/\psi$ near
$T_c$~\cite{Morita:2007pt,Gubler:2010cf,Gubler:2011ua}. The advantage of
this approach is that the temperature dependence of the operator
product expansion (OPE) for the charmonium current correlator can
be reliably obtained near $T_c$~\cite{Morita:2007pt,Kim:2015xna} through
 lattice calculations of the energy momentum
tensor. Unfortunately, in contrast to the vacuum
case, the $J/\psi$ pole in the spectral density will acquire a thermal
width above $T_c$ so that the changes in the OPE can either
be related to a mass change, an increase in the
width~\cite{Leupold:1997dg}, or a combination of these effects.
Nevertheless, it was shown that this approach
leads to a constraint for the changes of mass $\delta m$ and width
$\Gamma$ for the  $J/\psi$ at each temperature, which can be
approximated as $\Gamma={\rm Constant}+\delta
m$~\cite{Morita:2007hv,Gubler:2020hft}.

It is important to note, however, that the overlap of the charmonium
current with the $J/\psi$ appears as the strength of the ground
state pole in the QCD sum rule approach and can be reliably
determined at each temperature. At the same time, the overlap strength
can be identified as the charmonium wave function at the spatial origin,
which is sensitive to the potential through the normalization condition.
Comparing the temperature dependencies of the overlap strength obtained
from QCD sum rules to those obtained by solving the
Schr\"odinger equation for a given potential, three of us were able to
show that the potential for the charmonium system at short distance
should be dominated by the free energy near $T_c$~\cite{Lee:2013dca},
while at larger separation distance the potential will have a fraction
of about 20\% of the internal energy~\cite{Gubler:2020hft}, hence called
a transitional potential.

Given the real part of the potential, one can add
an effective imaginary part that is composed of
the leading order perturbative form derived in Ref.~\cite{Laine:2006ns},
multiplied by a free parameter $K$.
Solving the Schr\"odinger equation with this potential, one obtains the corresponding real and imaginary eigenvalues, which can be related to $\delta m$ and $\Gamma$ for each value of K. This leads
to another constraint equation involving these two quantities near $T_c$.

By combining the abovementioned two independent
constraints among $\delta m$ and $\Gamma$,
it becomes possible to  identify the mass shift and thermal width
separately for the $J/\psi$ near $T_c$, which is known to be a
non-perturbative and strongly interacting region. Furthermore, the result
 allows us to
quantify the strength of the imaginary potential for the heavy quark system near $T_c$.

{\it $J/\psi$ from heavy quark potential:}
The strong interaction between a heavy quark and
its anti-quark in our approach is modeled as a combination of the
free energy obtained from lattice QCD
calculations~\cite{Karsch:1987pv,Satz:2005hx,Gubler:2020hft} and the
internal energy derived from thermodynamic relations. Three of us
recently found that the long distance behavior of the heavy quark real
potential is composed of 80\% of the
free energy and 20 \% of the internal energy in order to reproduce
the $D$ meson mass near $T_c$ from QCD sum
rules~\cite{Gubler:2020hft}, while the short distance behavior
should be close to the free energy
potential~\cite{Lee:2013dca}.
The behavior of such a potential can be parametrized as

\begin{eqnarray}
V_R(r,T)=V_S(r,T)+\{V_L(r,T)-V_S(r,T)\}\nonumber\\
\times\frac{\tanh[(r-r_0)/\delta]+1}{2},
\end{eqnarray}
where $V_S(r,T)=F(r,T)$ and $V_L(r,T)=0.8~ F(r,T)+0.2~ U(r,T)$ are the short and long distance potentials, respectively. The
potential transits from $V_S$ to $V_L$ at $r_0=1~{\rm fm}$ with the width $\delta=0.25~{\rm fm}$.  Changing $r_0$ and $\delta$ within our constraints on the overlap of the wave function at the origin, we find that our final result on the dissociation temperature do not change significantly.
We adopt the imaginary heavy quark potential $V_I(r,T)$ calculated in HTL resumed perturbation theory~\cite{Laine:2006ns}:

\begin{eqnarray}
V_I(r,T)=-i\frac{g^2TC_F}{4\pi}\phi(m_Dr),
\label{laine}
\end{eqnarray}
where $g^2$ and $m_D$ are respectively the strong coupling
constant squared and Debye mass parametrized in~\cite{Kajantie:1997tt}, and
%\begin{eqnarray}
%g^2=\frac{8\pi^2}{9\ln (9.082 T/\Lambda_{\overline{MS}})},~~~m_D^2=\frac{4\pi^2T^2}{3\ln(7.547T/\Lambda_{\overline{MS}})},\nonumber
%\end{eqnarray}
%with $\Lambda_{\overline{MS}}=$300 MeV, and
\begin{eqnarray}
\phi(x)=2\int_0^\infty \frac{dzz}{(z^2+1)^2}\bigg[1-\frac{\sin(zx)}{zx}\bigg],
\end{eqnarray}
which has the limiting values $\phi(0)=0$ and $\phi(\infty)=1$.
We have taken $V(r,T)=V_R(r,t)+i KV_I(r,t)$, where the
multiplicative constant $K$ is varied from 0 to 4.
The heavy quark potential is then used in the Schr\"{o}dinger
equation of a charm and anti-charm pair as
\begin{eqnarray}
\bigg[2m_c-\frac{1}{m_c}\nabla^2+V(r,T)\bigg]\psi(r,T)=M^{J/\psi}(T) \psi(r,T),\label{schrodinger1}
\end{eqnarray}
where $m_c=1.25$ GeV is the bare charm quark mass, $\psi(r,T)=\psi_R(r,T)+i\psi_I(r,T)$
and $M^{J/\psi}(T)=M_R^{J/\psi}(T)+iM_I^{J/\psi}(T)$ the charmonium wave function and mass, respectively, at temperature $T$.

\begin{figure}[h]
\centerline{
\includegraphics[width=9 cm]{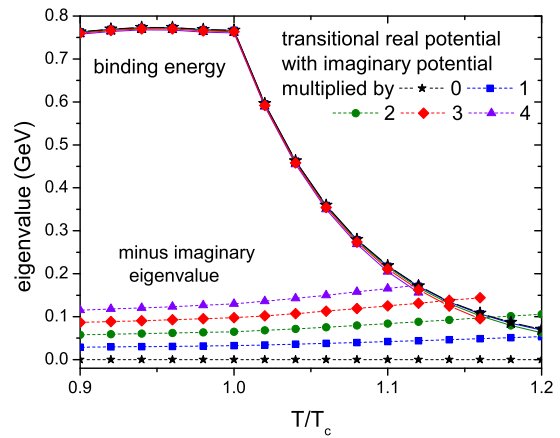}}
\caption{(Color online) Binding energies and imaginary eigenvalues of
 $J/\psi$ as a function of temperature for several values
 of $K$ multiplied to the imaginary potential of Eq.~\eqref{laine}~\cite{Laine:2006ns}.}
\label{eigenvalues}
\end{figure}

\begin{figure}[h]
\centerline{
\includegraphics[width=9 cm]{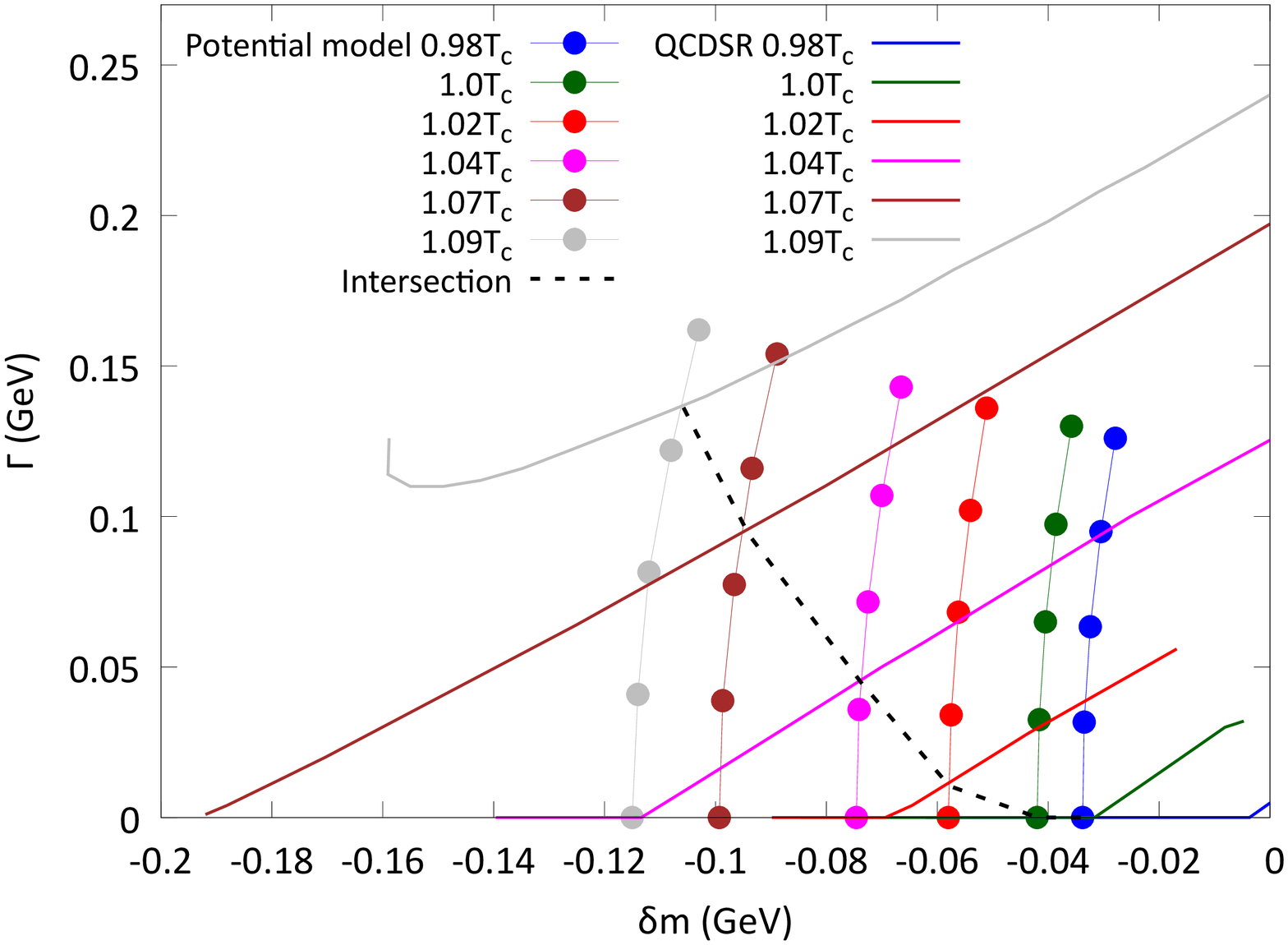}}
\caption{(Color online) The constraints between $\delta m$ and $\Gamma$ obtained from QCD sum rules (thick solid lines) and
the potential model (thin lines with filled circles). The filled circles in each thin line correspond to $K$ factors starting from 0 (bottom) to 4 (top).}
\label{deltam-gamma}
\end{figure}

Fig.~\ref{eigenvalues} shows the binding energy $\epsilon_R=2m_c+V_R( \infty, T)-M^{J/\psi}_R$  and imaginary eigenvalues obtained by numerically solving the coupled real and imaginary parts of Eq.~(\ref{schrodinger1}) with $K$ integers shown by different symbols.
As one can see, the binding energies  have only weak dependence on the strength of the imaginary
potential, while the imaginary eigenvalue scales with the
multiplicative strength of the imaginary potential.
Assuming that the strength of the imaginary potential is not known, we can obtain a constraint equation between the mass shift $\delta m= M_R^{J/\psi} (T) -M_R^{J/\psi}(0)$ and the thermal width $\Gamma= M_I^{J/\psi}(T)$ at a given temperature.
The results are shown as thin lines with filled circles in Fig.~\ref{deltam-gamma}.

{\it $J/\psi$ from QCD sum rules:}
The details of the sum rule calculations of
this work are adapted from Ref.\,\cite{Morita:2009qk}.
One starts with the Borel transformed energy dispersion relation for the invariant part of the correlator of the operator $j^{\mu}(x) = \bar{c}(x) \gamma^{\mu} c(x)$,
\begin{eqnarray}
\widehat{\Pi}(M^2, T) = \int_{0}^{\infty} ds e^{-s/M^2} \rho(s, T),
\label{sum_rules}
\end{eqnarray}
where $\widehat{\Pi}(M^2, T)$ represents the Borel transformed operator product expansion (OPE) of the correlator and $M$ stands for the Borel mass.  $\rho(s, T)$ is the vector channel spectral function at temperature $T$.
For the charmonium sun rule in medium,
the OPE is well determined by the temperature dependencies of dimension 4 gluon operators through lattice calculations~\cite{Boyd:1996bx,Kaczmarek:2004gv} of the energy density $\epsilon(T)$ and pressure $p(T)$~\cite{Morita:2007hv}.
Such a description of the OPE is valid up to temperatures slightly above $T_c$, where the temperature corrections are smaller than the vacuum values~\cite{Morita:2007pt} and the contribution from higher dimensional operators are small~\cite{Kim:2015xna}.

The behavior of the spectral function is assumed to have the following pole and continuum contribution
\begin{eqnarray}
\rho^{\mathrm{pole}}(s, T) &=& \frac{1}{\pi}
\frac{f \Gamma \sqrt{s}}{(s - m^2)^2 + s\Gamma^2}, \, s>4m_c^2, \label{pheno_side_pole}
\\
\rho^{\mathrm{cont}}(s, T) &=& \frac{1}{\pi} \theta(s - s_0) \mathrm{Im} \tilde{\Pi}^{\mathrm{pert}}(s),
\label{pheno_side_2}
\end{eqnarray}
where
the perturbative spectral function $\frac{1}{\pi} \mathrm{Im} \tilde{\Pi}^{\mathrm{pert}}(s)$ is given for instance
in Appendix B of Ref.\,\cite{Morita:2009qk}.

To eliminate the dependence on the residue $f$, which itself was critical in determining the potential at short
distance \cite{Lee:2013dca},
we consider the ratio
\begin{eqnarray}
&& \frac{-\frac{\partial}{\partial(1/M^2)}\bigl[ \widehat{\Pi}(M^2, T) - \widehat{\Pi}^{\mathrm{cont}}(M^2, T) \bigr]}{\widehat{\Pi}(M^2, T) - \widehat{\Pi}^{\mathrm{cont}}(M^2, T)} \nonumber \\
&=& \frac{\int_{4m_c^2}^{\infty}ds s e^{-s/M^2 } \rho^{\mathrm{pole}}(s, T) }{\int_{4m_c^2}^{\infty}ds e^{-s/M^2 } \rho^{\mathrm{pole}}(s, T)},
\label{effective_mass}
\end{eqnarray}
with
\begin{eqnarray}
\widehat{\Pi}^{\mathrm{cont}}(M^2, T) = \int_{0}^{\infty} ds e^{-s/M^2} \rho^{\mathrm{cont}}(s, T).
\label{effective_mass_2}
\end{eqnarray}
For given $s_0$ and $\Gamma$, Eq.~(\ref{effective_mass}) can be solved for the $m(M^2,T)$, appearing in Eq.~(\ref{pheno_side_pole}), as a function of the Borel mass $M^2$.
The admissible range of $M^2$ is fixed by the so-called Borel window.
The minimum value $M^2_{\mathrm{min}}$ is determined by the condition that the dimension 4 OPE terms are smaller than 30\,\% of the
total OPE expression, the maximum
$M^2_{\mathrm{max}}$ from the condition that
the continuum term  contributes less than 30\,\% of the perturbative term to the sum rule.

Next, we define
\begin{eqnarray}
\chi^2 &\equiv& \frac{1}{M^2_{\mathrm{max}} - M^2_{\mathrm{min}}} \nonumber \\
&&\times \int_{M^2_{\mathrm{min}}}^{M^2_{\mathrm{max}}} dM^2 \bigl[ m(M^2,T) - m(M_0^2,T) \bigr]^2,
\label{effective_mass_3}
\end{eqnarray}
where $M^2_0$ is the value of $M^2$, at which the derivative of $m(M^2,T)$ by $M^2$ vanishes: $dm(M^2,T)/dM^2\bigr|_{M^2 = M^2_0} = 0$.
$m$ is determined as $m(M^2_0,T)$ and $\Gamma$ is chosen such that $\chi^2$ is minimal.
Hence, starting from three independent parameters $m,\Gamma,s_0$, Eq.~(\ref{effective_mass}) and  Eq.~(\ref{effective_mass_3}) effectively lead  to a constraint among $\delta m$ and $\Gamma$.
The obtained constraint relations~\cite{Morita:2009qk} are shown as solid thick lines without symbols in Fig.~\ref{deltam-gamma}.

{\it Matching the heavy quark potential to QCD sum rules:}

\begin{figure}[h]
\centerline{
\includegraphics[width=8 cm]{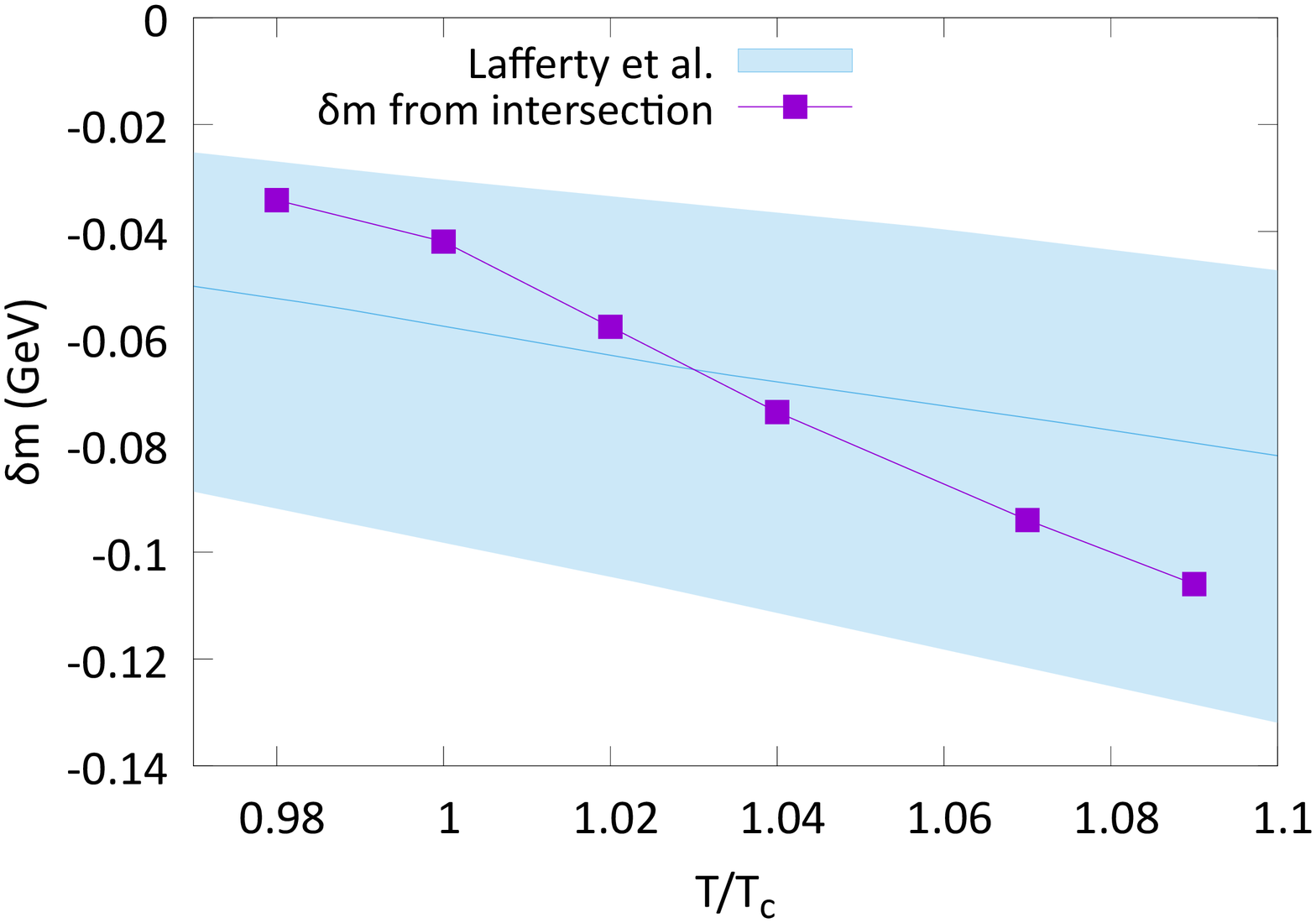}}
%\caption{(Color online) Temperature dependence of $J/\psi$ mass.}
%\label{deltam-T}
%\end{figure}
%\begin{figure}[h]
\centerline{
\includegraphics[width=8 cm]{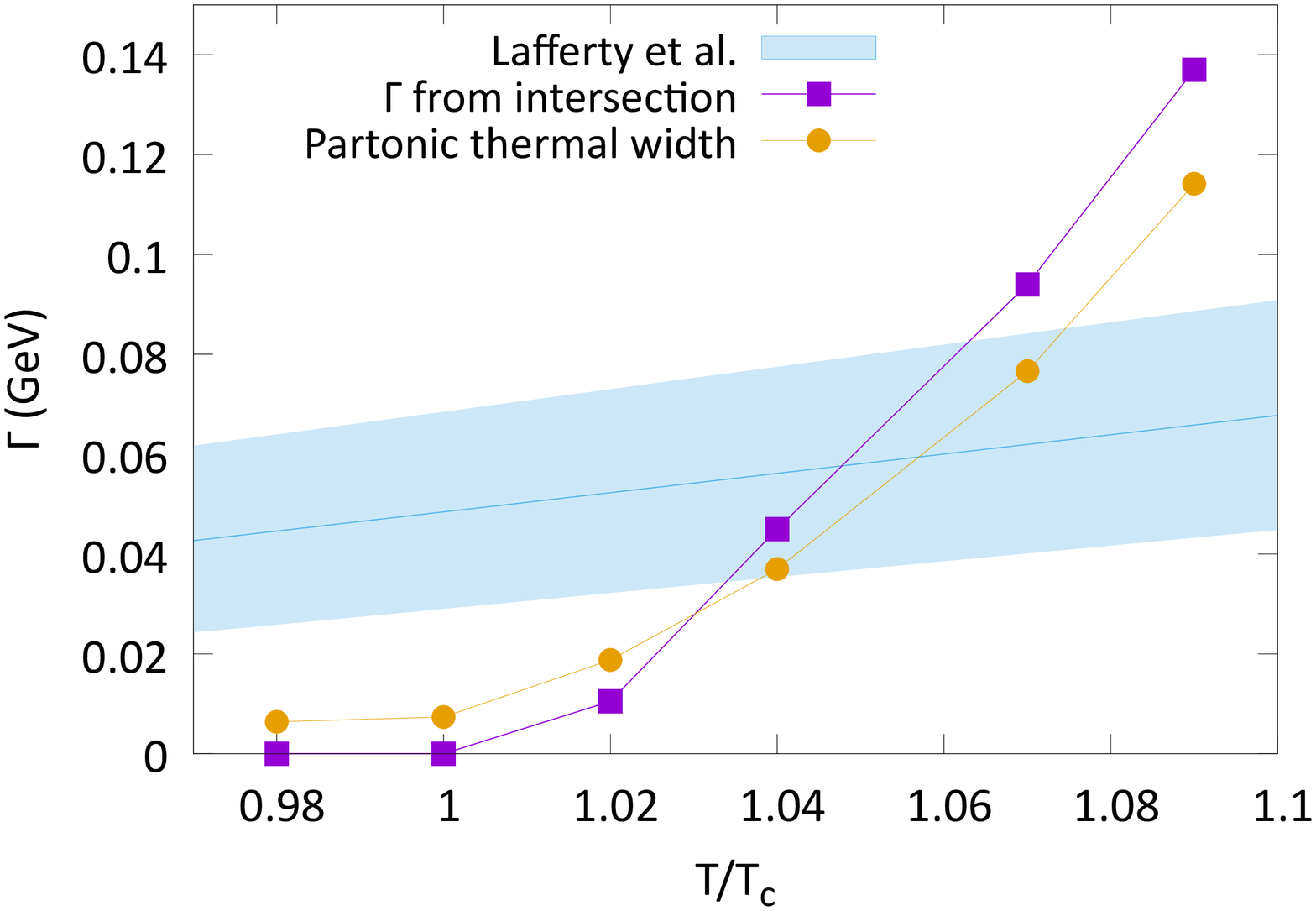}}
\caption{(Color online) Temperature dependence of $J/\psi$ mass and width.
The blue lines and surrounding shaded regions are adapted from Ref.~[29]. The violet (orange)
lines with filled rectangles (circles) are the results of this work, obtained from the intersection
points of Fig.\,\ref{deltam-gamma} (partonic HTL resummation).}
\label{gamma-T}
\end{figure}

Let us now study the consequences of combining the two
independent constraints for $\delta m$ and $\Gamma$. In
Fig.~\ref{gamma-T}, we respectively plot $\delta m$ and $\Gamma$ from
the intersection points of the two constraints given in
Fig.~\ref{deltam-gamma}. One notes that $|\delta m| \sim \Gamma$
at $T=1.08~ T_c$, a bit below the highest temperature considered
in this work.
Also plotted in Fig.~\ref{gamma-T} are results from a recent extraction of $\delta m$ and $\Gamma$ from lattice QCD data~
\cite{Lafferty:2019jpr}.  One notes that the temperature dependence of $\delta m$ are within the bounds of the lattice results.
For $\Gamma$, our results show a stronger temperature dependence
than those of Ref.\,\cite{Lafferty:2019jpr}, with a comparable magnitude
just above $T_c$.

From the intersection points, we can moreover determine the strength of the imaginary potential. As can be seen  in Fig.~\ref{deltam-gamma}, the multiplicative constant $K$ starts from 0 at $T=0.98 ~T_c$ and
rises to about 4 at $T=1.09 ~T_c$, demonstrating the highly non-perturbative behaviour near $T_c$.

Combining the above observations and Fig.~\ref{eigenvalues}, one sees that the thermal width
of the $J/\psi$ exceeds the binding energy above $1.1T_c$, suggesting
that that the dissociation will occur there because the $J/\psi$ will lose its identity when the width becomes
sufficiently larger than the binding energy \cite{Mocsy:2007jz}. This is
consistent with the fact that the dissociation will occur around
1.1~$T_c$ when the free energy potential is used.

Within the potential nonrelativistic QCD (pNRQCD) approach, related transport coefficients can be obtained through the relation $\delta m(1S)=\frac{3}{2} a_0^2 \gamma$ and $\Gamma(1S)=3a_0^2 \kappa$~\cite{Brambilla:2019tpt}, where $\kappa$ is the diffusion coefficient of a heavy quark in the medium, $\gamma$ its dispersive counterpart and  $a_0$  the Bohr radius.
Assuming $a_0=0.23$~fm and $T_c=155 {\rm ~MeV}$, we find that our data yield $ -10.78 \leq  \gamma/T^3 \leq -5.63$  and $ 0 \leq \kappa/T^3 \leq  6.97$ for $T_c \leq T \leq 1.09~T_c$,
which is close to the findings of several methods summarized in Ref.~\cite{Brambilla:2019tpt}.

In Fig.~\ref{gamma-T}, we also plot the thermal width for $J/\psi$ using a formula recently derived by a
partonic description that includes the HTL resummation and  reduces to low energy theorems obtained by
pNRQCD in the relevant kinematical limit~\cite{Brambilla:2013dpa} so that it can be used in
a wide temperature range applicable to heavy quark systems in heavy ion collisions~\cite{Hong:2018vgp}. The inputs in the calculations are the effective Bohr radius
($a_0^2=\frac{1}{3}\int \frac{d^3p}{(2\pi)^3} \, |\nabla\psi(p)|^2$) and
the binding energy, for which we used the results from the present potential model calculation, and the two-loop perturbative running coupling constant in Ref.~\cite{Morita:2007hv}.
The partonic thermal width exhibits a behavior that is consistent with the results of this work.

{\it Summary and Conclusions:}
Combining the two constraint equations for $\delta m$ and $\Gamma$ discussed in this work, we are now finally able to separately identify the temperature dependencies of these observables near $T_c$, showing that the $J/\psi$ will melt slightly above $T_c$.
We also identified the strength of the imaginary potential for the heavy quark- anti-quark system near $T_c$, which is found to have a large multiplicative factor of about 4 relative to the leading perturbative results at  1.09  $T_c$, which lies in the highly non-perturbative temperature domain. Our results are within the limits of a recent lattice calculation \cite{Lafferty:2019jpr},
and the  thermal width is close to that obtained by using a partonic picture with resummed pQCD \cite{Hong:2018vgp}.
They are based on stringent constraints from two independent methods and form reliable nonperturbative results for the properties of $J/\psi$ as well as the real and imaginary potential near $T_c$ and will thus provide valuable input to our understanding of the heavy quark system in heavy ion collisions.

\section*{Acknowledgements}
The authors acknowledge useful discussions with N.~Brambilla and A.~Vairo, which partly motivated the work.
This work was supported by the Samsung Science and Technology Foundation under Project Number SSTF-BA1901-04.
P.G. is supported by Grant-in-Aid for Early-Carrier Scientists Nos. JP18K13542 and JP20K03940, and the Leading Initiative for
Excellent Young Researchers (LEADER) of the Japan
Society for the Promotion of Science (JSPS).

\end{document}